\documentstyle[editedbook,epsfig,psfig,epsf]{mq}

\def\cm2{cm$^2$ }
\def\se1{s$^{-1}$ }

\begin{opening}
\title{Fast Infrared Sub-Flares From GRS 1915+105}
\author{D.M. Rothstein$^1$ \& S.S. Eikenberry$^1$} \institute{$^1$
Department of Astronomy, Cornell University, Ithaca, NY, USA 14853}
\end{opening}

\runningtitle{IR Sub-Flares From GRS 1915+105}
\runningauthor{Rothstein \& Eikenberry}

\begin{document}
\vspace{-0.5cm}
\begin{abstract}
{\small We report the discovery of fast ($\sim 100$ sec) 2.2 $\mu${m}
flares from the microquasar GRS 1915+105, which are superimposed on
longer ($\sim 30$ min), brighter flares corresponding to episodes of
jet formation.  The number and strength of the sub-flares in each
bright flare varies, and does not seem to correlate in any obvious way
with the underlying light curve or with changes in the X-ray emission.
However, the fact that these sub-flares only occur in tandem with the
larger flares indicates that they might be related to the jet ejection
process.}
\end{abstract}

\section{Introduction}

\begin{figure}[htb]
\centering \epsfig{file=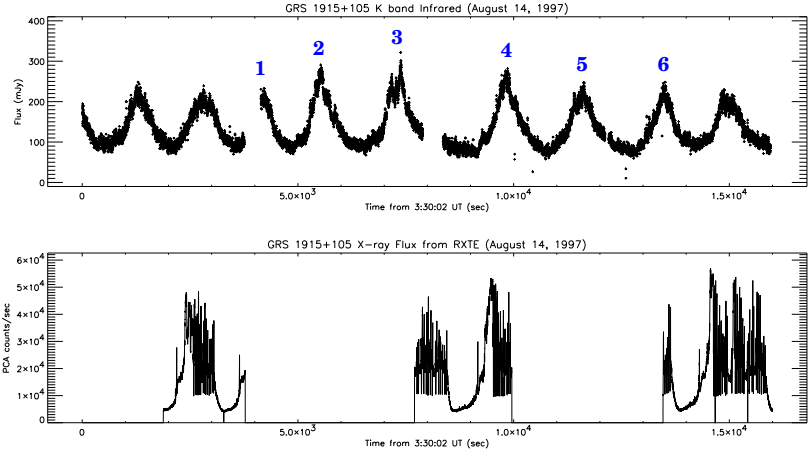,width=11cm}
\caption{Light curves of GRS 1915+105 on August 14, 1997 UT (Eikenberry
et al. 1998).
The data in the top panel were taken in K-band (2.2
microns) using the Palomar 5-meter telescope and have been dereddened
by 3.3 magnitudes.  The data in the bottom panel were taken using the
Proportional Counter Array (PCA) on the Rossi X-ray Timing Explorer
(RXTE).  Both light curves have time resolution of one second.  The
numbered infrared flares correspond to those shown in Figure 2.}
\label{fig:lcurve}
\vspace{-0.2cm}
\end{figure}

Simultaneous X-ray and infrared light curves of the microquasar GRS
1915+105 were obtained on August 14, 1997 \cite{eiken}, one of the
first observations showing the disk-jet interaction in this source
(Figure \ref{fig:lcurve}).  GRS 1915+105 exhibited quasi-regular
flaring behavior in the infrared during the course of these
observations (interpreted as synchrotron emission from ejected
plasma), and the flares were seen to correlate with changes in the
X-ray emission (interpreted as emptying and refilling of the system's
inner accretion disk).\\ \indent Smaller amplitude variability, on
faster timescales, was also observed in some of the infrared flares.
In this contribution, we have reanalyzed the original data (obtaining
higher signal-to-noise photometry than previously presented) and
identified all such ``sub-flares'' above a given detection threshold,
in an attempt to characterize this previously unknown phenomenon.

\begin{figure}[htb]
\centering
\psfig{file=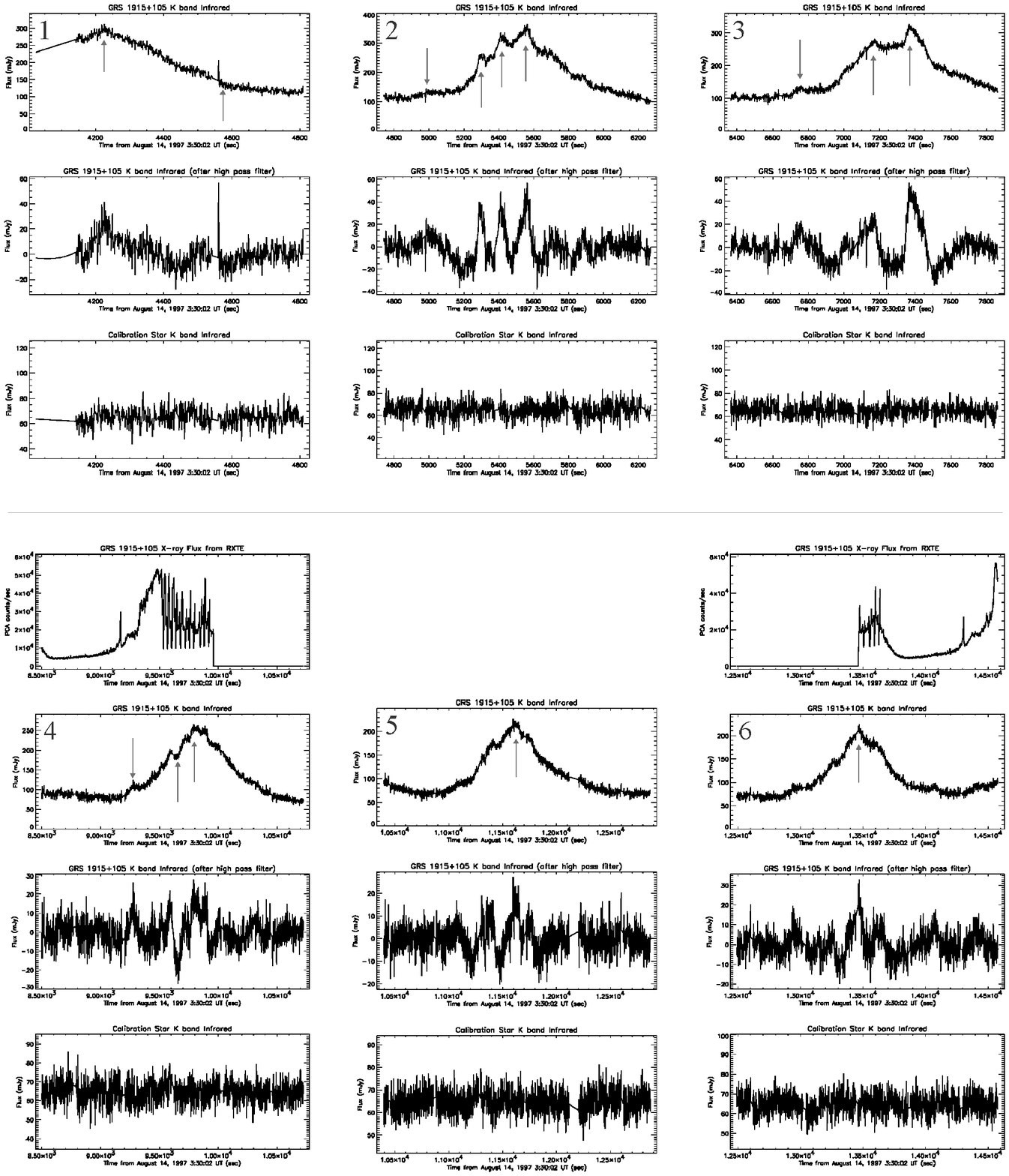,width=17cm}
\caption{The reanalyzed data from Figure 1, showing the
six infrared flares in which sub-flares were detected.  In each case,
the numbered panel shows the reanalyzed light curve from Figure
1, with arrows marking each statistically significant
sub-flare.  The next panel down shows the light curve after
application of a high pass filter to remove all frequencies lower than
$\sim$ 1/(500 sec).  The bottom panel shows the light curve of a field
star near GRS 1915+105, which we used to calibrate variation due to
atmospheric effects.  In (4) and (6), the top panel shows the X-ray
light curve from Figure 1 at the time corresponding to
the sub-flaring behavior.}
\label{fig:sub}
\vspace{-0.25cm}
\end{figure}

\section{Data Analysis}
The infrared data presented here consist of observations of GRS
1915+105 and a nearby field star, which we used for calibration
purposes.  We detected sub-flares using the following procedure:
\begin{itemize}
\item At each time step, we added a $2\%$ systematic error to the
measurement errors from the photometry.  This was chosen in order to
make the errors for the calibration star roughly match those
determined from the variability in its light curve.
\item We then ran the GRS 1915+105 and calibration star light curves
through a high pass filter to remove all frequencies lower than $\sim$
1/(500 sec).
\item We searched for sub-flares centered at each point in the light
curve on timescales ranging from 25 to 200 seconds and identified
those with reduced chi-squared above a threshold value.  The threshold
was chosen to make the probability of obtaining one false detection
across the entire light curve $< \sim 1\%$, and also to ensure that
any atmospheric variability too weak to observe in the (fainter)
calibration star did not lead to any false detections of sub-flares in
GRS 1915+105.
\item We eliminated GRS 1915+105 sub-flares that correlated in time
with calibration star flares, assuming these were due to atmospheric
variability.
\end{itemize}

\section{Discussion}
The sub-flares we detected are shown in Figure \ref{fig:sub}.  They
have typical rise times of 40 to 100 seconds (with a few longer) and
typical amplitudes of 25 to 70 mJy.  The most important point about
the sub-flares is that they are all superimposed on larger flares,
indicating a possible relationship to the jet.  Some of the large
flares are multiple-peaked (2,3), with several strong sub-flares
superimposed on them, while other large flares only have one peak
(1,6), but it is sharp enough to be detected as a sub-flare by our
algorithm.\\
\indent There is no obvious correlation between the properties of the
main flare and the number and strength of sub-flares superimposed on
it.  The X-ray coverage of the sub-flare detections is poor, but in
one case for which they overlap (4), a faint sub-flare is detected at
the beginning of the main flare, indicating a possible connection to
the X-ray ``spike'' which marks the beginning of the jet ejection.
Two other flares (2,3) also show evidence for a faint sub-flare near
the beginning of the main flare.  Taken together, this suggests that
these sub-flares originate near the accretion disk and might be
associated with the jet formation process.

\section*{Acknowledgments}
DMR is supported in part by a National Science Foundation Graduate
Research Fellowship.  SSE and DMR are supported in part by an NSF
CAREER award (NSF-9983830).

\end{document}